\documentclass[https://www.overleaf.com/project/5daf17bb9a3ada00015dd4eaaps,prfluids,superscriptaddress,amsfonts,amssymb,amsmath,showpacs,longbibliography]{revtex4}

\usepackage{amsmath,amssymb}
\usepackage[T1]{fontenc}
\usepackage[utf8]{inputenc} 
\usepackage{graphicx}
\usepackage[dvips]{epsfig}
\usepackage{subfigure}
\usepackage{xcolor}
\usepackage{color}
\usepackage{float}
\usepackage{lipsum}
\usepackage{microtype}
\usepackage{siunitx}

\usepackage{ulem}
\usepackage{enumitem}
\usepackage[colorlinks=true, linkcolor=royal, urlcolor=royal, citecolor=royal, anchorcolor=royal]{hyperref}

\usepackage{url}
\usepackage{comment}
\usepackage{dcolumn}
\usepackage{lmodern}
\usepackage{bm}
\usepackage{graphicx,subfigure,tikz,pgfplots,comment}

\definecolor{royal}{RGB}{0,0,50}
\definecolor{greenSimon}{RGB}{20,150,40}
\usepackage[colorinlistoftodos]{todonotes}

\begin{document}


\title{Mixing and unmixing induced by active camphor particles}

\author{Cl\'ement Gouiller}
\affiliation{ILM, Univ Lyon, Univ Lyon 1, CNRS, F-69622 Villeurbanne CEDEX, France}

\author{Florence Raynal}
\affiliation{LMFA, Univ Lyon, Centrale Lyon, INSA Lyon, Univ Lyon 1, CNRS, F-69134 \'Ecully, France}

\author{Laurent Maquet}
\affiliation{ILM, Univ Lyon, Univ Lyon 1, CNRS, F-69622 Villeurbanne CEDEX, France}

\author{{Micka\"el Bourgoin}}
\affiliation{Laboratoire de Physique, Univ Lyon, ENS de Lyon, Univ Lyon 1, CNRS, F-69342 Lyon, France}

\author{C\'ecile Cottin-Bizonne}
\affiliation{ILM, Univ Lyon, Univ Lyon 1, CNRS, F-69622 Villeurbanne CEDEX, France}

\author{Romain Volk}
\affiliation{Laboratoire de Physique, Univ Lyon, ENS de Lyon, Univ Lyon 1, CNRS, F-69342 Lyon, France}

\author{{Christophe Ybert}}
\affiliation{ILM, Univ Lyon, Univ Lyon 1, CNRS, F-69622 Villeurbanne CEDEX, France}
\date{\today}



\date{\today}

\begin{abstract}
In this experimental study, we report on the mixing properties of interfacial colloidal floaters (glass bubbles) by chemical and hydrodynamical currents generated by self-propelled camphor disks swimming at the air-water interface. 
Despite reaching a statistically stationary state for the glass bubbles distribution, those floaters always remain only partially mixed. 
This intermediate state results from a competition between (i) the mixing induced by the disordered motion of many camphor swimmers and (ii) the unmixing promoted by the chemical cloud attached to each individual self-propelled disk. 
Mixing/unmixing is characterized globally using  the standard deviation of concentration and spectra, but also more locally by averaging the concentration field around a swimmer. 
Besides the demixing process, the system develops a "turbulent-like" concentration spectra, with a large-scale region, an inertial regime and a Batchelor region.
We show that unmixing is due to the Marangoni flow around the camphor swimmers, and is associated to compressible effects.
\end{abstract}

\maketitle

\section{Introduction}
Particles floating on water and stirred by turbulence tend to segregate into string-like clusters~\cite{bib:cressman2004_NJP, bib:larkin2009_PRE,bib:lovecchio2013_PRE,bib:lovecchio2014} as they disperse. 
Similarly, a passive scalar (let say the concentration field of a given substance) stirred on a turbulent interface is also found to exhibit pronounced persistent heterogeneities~\cite{bib:eckhardt2001_PRE}, never reaching a perfectly mixed state. 
Such a \textit{turbulent interfacial unmixing} can have dramatic consequences in environmental, geophysical and industrial flows. 
It is for instance responsible for the accumulation of pollutants (such as micro-plastic debris) floating at the surface of the oceans \cite{bib:lebreton2012_MarinePollutionBull,bib:eriksen2014_PlosOne}, and is probably also to be associated to coastal mixing fronts~\cite{bib:carrillo2001_PhysChemEarth} as well as to the dynamics of phyto-plankton blooms and patchiness~\cite{bib:piontkovski1997,bib:abraham2000_Nature}.

In this respect, clustering of floaters on a turbulent interface contrasts with the usual intuition that turbulence always enhances the mixing of tracer particles and passive scalars, with an effective turbulent diffusivity orders of magnitude larger than the simple molecular diffusivity, classically leading to a rapid homogenization of stirred substances.
This is all the more true that such a clustering  is observed regardless of any attractive interactions between floaters, such as the so-called \textit{cheerios effect}~\cite{Vella:2005hi}.
Indeed, such counter-intuitive response to turbulent stirring is known to occur when there exist sources of compressibility in the system. This has been long recognized for the transport of inertial particles, with the so-called preferential concentration phenomenon~\cite{bib:monchaux2012_IJMF} associated to the effective compressibility that arises from inertia-induced departure between particles' velocity and stirring -- incompressible-- flow velocity~\cite{bib:maxey1987}).

In the case of floaters, the compressibility source responsible for the persistent clustering does not rely on additional physical effects as above, but is readily provided by the flow. Indeed the 2D interfacial turbulent field of relevance is a priori compressible due to the presence of upwelling regions (acting as divergent zones or sources) and downwelling regions (acting as convergent zones or sinks), even in the case where the 3D stirring turbulence underneath the interface is itself incompressible~\cite{bib:eckhardt2001_PRE, bib:larkin2009_PRE,bib:afonso2009_JPA,bib:lovecchio2013_PRE}.
This results in a  final state which is statistically stationary (though out-of-equilibrium) where on one hand turbulent mixing tends to homogenize the particles spatial distribution while compressibility effects sustain the non-uniformities with dense regions at the convergent zones and depleted regions at the divergent zones.

For both the turbulent clustering of floaters and the one of inertial particles the effective compressibility is related to some underlying coupling to the overall turbulent mixing process.
In recent studies~\cite{bib:Volketal2014,bib:Maugeretal2016,bib:shukla2017_NJP,bib:raynaletal2018,bib:raynal_volk2019} our groups have shown that similar compressible effects on the transport and mixing properties of particles can be induced by the particle response -- a so-called phoretic drift -- to environmental field gradients.
This \textit{environment sensing} strategy can be built from various phoretic phenomena: diffusiophoresis (drift induced by chemical concentration gradients), thermophoresis (drift induced by thermal gradients), electrophoresis (drift induced by electric field gradients), etc. 
Experiments, simulations and analytical models~\cite{bib:Volketal2014,bib:Maugeretal2016,bib:shukla2017_NJP,bib:raynaletal2018, bib:raynal_volk2019} show that such phoretic particles acquire an effectively compressible dynamics, even though the underlying flow is perfectly incompressible.
Active (self-propelled) particles, which can be seen as an extreme case of phoretic particles (with self-generated gradients driving their drift), have also been reported to exhibit clustering when stirred by incompressible chaotic or turbulent flows~\cite{bib:khurana2011_PRL, bib:durham2011_PRL}.

In this article, we propose to experimentally investigate a new original configuration of interfacial mixing, where small floating particles are stirred by active interfacial particles, in the absence of any other underlying forced flow. 
More precisely, we consider the mixing of a patch of micrometric hollow glass spheres floating on water, stirred by millimetric interfacial active camphor disks~\cite{bib:bourgoin_etal2020}. 
The motivation of exploring this configuration is two-folds: (i) in a recent study~\cite{bib:bourgoin_etal2020} we have shown that the dynamics of such active disks mimics very accurately the statistical multi-scale properties of homogeneous isotropic turbulence (in particular a Kolmogorov like spectrum has been found), although the underlying flow itself remains almost at rest; it is therefore tempting to investigate the mixing induced when such active disks are used as stirrers. 
(ii) The self-motility of these camphor disks is driven by self generated surface tension gradients (related to a symmetry breaking of the dissolution of camphor in water) \cite{Nakata:2015ki}; one may therefore expect that the mixed micro-particles experience some additional phoretic drift (due to the surface tension gradients) eventually leading to an effective compressibility behaviour, as previously described. As a consequence, although no actual fluid turbulence (neither compressible nor incompressible) is present in the system, the proposed configuration shares qualitative properties with interfacial turbulent mixing: a stirring mechanism (active camphor particles) with turbulent-like statistical features and a source of possible compressible effects (related to the presence of surface tension gradients).

We explore the global and multi-scale mixing properties when a small patch of glass bubbles is released at the free surface and stirred by the active camphor disks. 
We observe, that similarly to interfacial turbulent mixing, a final non-uniform steady state is reached for the concentration field of the micro-particles, with the existence of densely seeded regions and depleted regions (in the trail of the active camphor stirrers). A spectral analysis of the concentration field reveals striking quantitative analogies with the turbulent mixing of a passive scalar. Finally a close investigation of the dynamics in the vicinity of individual stirrers confirms the role played by compressible effects induced by Marangoni flows driven by tension surface gradients in the chemical wake of the active camphor stirrers.

\section{Materials and Methods}
\subsection{Experimental setup}
\begin{figure}
    \centering
\includegraphics{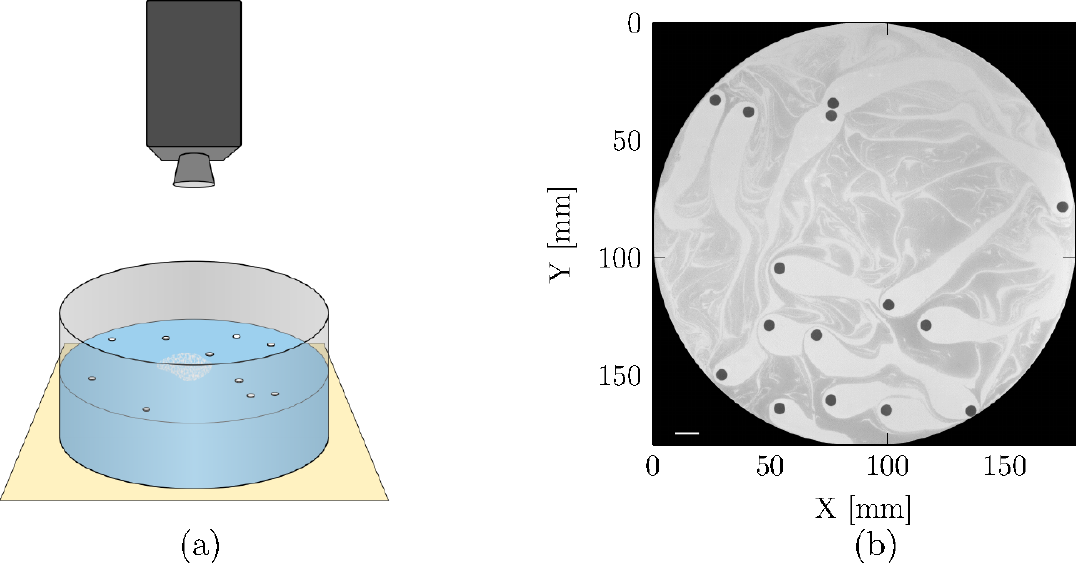}
      \caption{(a) Experimental setup: water tank filled with \SI{1}{cm} millipore water. 
      At the surface, $N$ interfacial swimmers (camphor disks). At the beginning of the experiment, a patch of passive floaters (\SI[parse-numbers = false]{40 \pm 2}{mg} of glass bubbles) is introduced in the system near the center of the bath.  A LED plate emits light from above and a camera acquires light transmitted through the setup. 
      (b) Typical image recorded after ten minutes for $N=15$ swimmers with radius $R=$\SI{2.5}{mm}.}
      \label{fig:setup}
\end{figure}

Camphor disk swimmers are made by punching an agarose sheet (\SI{0.5}{mm} in thickness) filled with solid camphor grains 
\cite{boniface2019self, Nakata:2015ki, bib:bourgoin_etal2020}. 
Depending on the size of the puncher, the radius $R$ of the camphor disks considered hereafter ranges from 1 to \SI{4}{mm}. 
A number $N$ (ranging from 7 to 45 \footnote{For small number of camphor disks (typically $N\le5$), swimmers have a tendency to follow the side of the container, and mixing is not efficient enough. In particular it is not possible to access to the averaged concentration field around a swimmer such as shown in figure \ref{fig:concentration}, because the glass bubbles do not cover enough domain. This is why we decided to begin with a minimum number of swimmers equal to 7.}) of freshly made swimmers is deposited at the surface of a \SI{1}{cm}-thick water subphase (Elga PureLab Flex ultra-pure water) in a circular glass cell (Fig. \ref{fig:setup}a).
All around the cell, a thin floating plastic ring is placed at the edge, that cancels the capillary meniscus, avoiding the trapping of floaters or swimmers; the available remaining free surface has a diameter of $\SI{18}{\centi\metre}$.
Prior to each new experiment, the subphase is renewed by fresh water.

Once deposited on water, swimmers begin to release camphor into the fluid which results in surface tension heterogeneities in their vicinity that drive the so-called Marangoni flows \cite{bib:Nishimori_jpsj-2017,bib:leRouxetal2016}.
Despite their circular shape, a spontaneous symmetry breaking occurs immediately, resulting in their propulsion in an arbitrary direction, at a typical velocity $U\sim \SI{6}{cm.s^{-1}}$ for a single swimmer of radius $R=\SI{2.5}{mm}$ \cite{boniface2019self}. 
This swimming velocity depends not only on the swimmers radius and  physico-chemical parameters, but also on the number of swimmers as they interact altogether.

As soon as enough swimmers are present, the system starts to exhibit spatio-temporal fluctuations \cite{bib:bourgoin_etal2020}, so that their velocity is better characterized using the root mean square $u_\mathit{rms}=\left(\langle u^2_x \rangle + \langle u^2_y\rangle\right)^{1/2}$.
Table \ref{table1} summarizes the values of $u_\mathit{rms}$ together with the corresponding Reynolds numbers, for the cases considered in our experiments; 
for a sake of comparison with other works, the percentage of surface fraction covered by the swimmers $\varphi_s=100\,N\pi R^2/A_t$, where $A_t$ is the total available area, is also given.
The Reynolds number is defined here as $\mathrm{Re}_p = u_\mathrm{rms} R/\nu$, where $\nu$ is the kinematic viscosity of water, and two series of experiments have been conducted. 
In the first case the radius is fixed ($R=\SI{2.5}{mm}$) and the number $N$ of Marangoni swimmers is increased, resulting in a decreasing velocity. 
In the second case we consider a fixed number of swimmers $N=15$, with increasing radius: here the rms-velocity does not vary much, although the Reynolds number increases with increasing radius.
Given the range of Reynolds numbers considered here, $Re_p \in [20,100]$, the flow of water remains perfectly laminar although the dynamics of the swimmers is fluctuating in space and time as observed in our previous study \cite{bib:bourgoin_etal2020}.

%
\begin{table}
\caption{Root mean square velocity of the swimmers $u_\mathit{rms}=\sqrt{\langle u^2_x \rangle + \langle u^2_y\rangle}$  and corresponding Reynolds number here defined as $\mathrm{Re}_p = u_\mathrm{rms} R/\nu$, where $\nu$ is the kinematic viscosity of water, for: 
(a) $R=\SI{2.5}{mm}$, and different numbers of swimmers $N$; (b) $N=15$, and increasing radius $R$. For all experiments we indicate $\varphi_s$, the percentage of surface fraction covered by the swimmers.}
  \centering
%

\begin{tabular}{|@{\quad}l | c |c |c |c |c |c |}
\multicolumn{6}{c}{$R = \SI{2.5}{\milli\meter}$}\\[.4mm]
\hline
$N$ & 7 & 11 & 15 & 20 & 30 & 45 \\[.4mm]
\hline
$\varphi_s$ & 0.54 & 0.85 & 1.16 & 1.54 & 2.32 & 3.47 \\[.4mm]
\hline
$u_\mathit{rms}$ (mm/s)\ \ &\ 37.1\  &\ 32.8\  &\ 26.6\  &\ 22.1\  &\ 15.8 \  &\ 11.2 \\[.4mm]
\hline
$\mathrm{Re}_p$ & 93 & 82 & 66 & 55 & 40 & 28 \\
\hline
\end{tabular}
\hfil
\begin{tabular}{|@{\quad}l | c |c |c |c |c |c |c |}
\multicolumn{6}{c}{$N = 15$}\\[.4mm]
\hline 
$R$ (mm) & 1 & 1.5 & 2 & 2.5 & 3 & 3.5 & 4 \\[.4mm]
\hline
$\varphi_s$ & 0.19 & 0.42 & 0.74 & 1.16 & 1.67 & 2.27 & 2.96 \\[.4mm]
\hline
$ u_{rms}$ (mm/s)\ \ &\ 19.4\ \ &\ 26.6\ \ &\ 26.7\ \ &\ 26.6\ \ &\ 25.1\ \ &\ 24\ \ &\ 22.8 \\[.4mm]
\hline

$\mathrm{Re}_p$ & 19 & 40 & 53 & 66 & 75 & 84 & 91\\
\hline
\end{tabular}\\
\vspace{0.1cm}
\quad(a)\hskip7.4cm(b)
\label{table1}
\end{table}
%
At the beginning of an experiment, a patch of passive floaters constituted of glass bubbles (see properties in \cite{glassbubbles}) is introduced in the system of interfacial swimmers near the center of the bath. 
The subsequent dynamics of the whole system, backlit with a LED panel, is recorded from top using a HXC Flare camera equipped with a Nikon 24-85mm f/2.8-4D IF AF NIKKOR objective, yielding images with resolution 2048x2048 px$^2$ at a rate of \SI{35}{Hz} (see fig. \ref{fig:setup}a). 
A typical image is shown in figure \ref{fig:setup}b where interfacial swimmers appear as dark disks and glass bubbles as grey shades on the surface.

\subsection{Concentration field}
\begin{figure}
\includegraphics{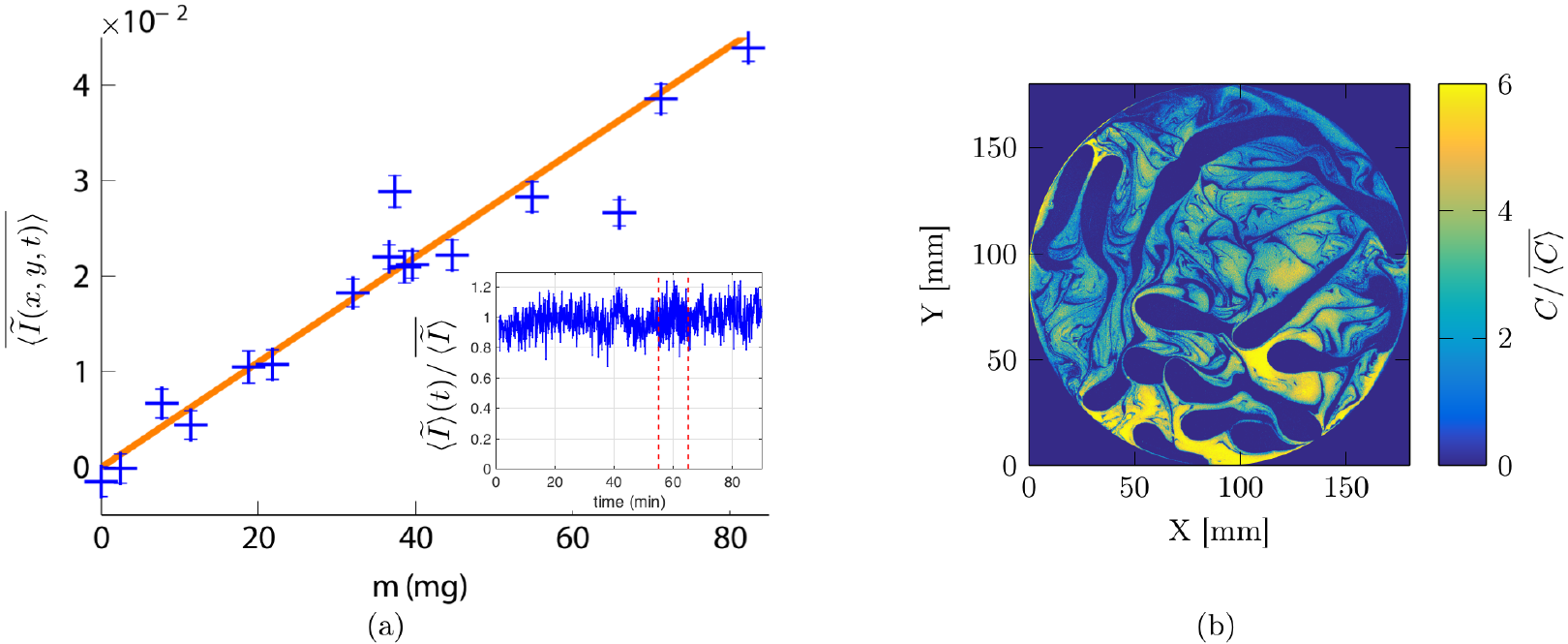}
      \caption{(a) Mean measured intensity, $\overline{\langle \widetilde I (x,y,t)\rangle}$, averaged over space and time, as a function of the poured glass bubbles mass $m$ for $N=15$ swimmers. $(+)$ measurements, $(-)$ linear fit $\overline{\langle \widetilde I \rangle} = a\times m$, with $a=5.51\times 10^{-4}\mathrm{mg}^{-1}$. The error bar is estimated as $\sigma_{\widetilde I}=10^{-3}$ by computing the standard deviation of the time series $\langle \widetilde I\rangle(t)$. The inset is a time series of the non dimensional instantaneous quantity $\langle \widetilde I\rangle(t)/\, \overline{\langle \widetilde I \rangle}$ as a function of time; the vertical dotted lines indicate our measurement window for time averaging. (b) Instantaneous non dimensional glass bubble concentration field deduced from the image Fig. \ref{fig:setup}b.}
      \label{fig:concentration}
\end{figure}

In the following, $\langle Q\rangle$ denotes the average of a given quantity $Q$ over the whole circular surface.
To access the local surface concentration of floaters, we quantify the light absorption at location $(x, y)$ and time $t$ by comparing the light intensity field $I(x,y,t)$ in presence of floaters to the intensity field $I_0(x,y)$ of a reference image without floaters. 
Assuming that such an absorption is linear with the local concentration $C(x,y,t)$ of the floaters --- a natural choice for a single layer of individual scatterers --- we define a non dimensional intensity field $\widetilde I(x, y,t)$ through the relation 
\begin{equation} 
\widetilde I(x, y,t) = \frac{I_0(x,y)-I(x,y,t)}{I_0(x,y)}. \label{calib} 
\end{equation} 
In order to support that this linear approximation gives a good indication of the local concentration of floaters, \textit{i.e.}
\begin{equation} 
\widetilde I(x, y,t) \propto C(x, y,t)\, , 
\label{I_propto_C}
\end{equation} 
we must check that $\langle\widetilde I(x,y,t)\rangle$ estimated from Eq.~(\ref{calib}) satisfies mass conservation and linearly follows the total number of glass bubbles. 
For a given amount of poured glass bubbles onto the surface, the total number of particles must indeed be conserved over time as the mixing process goes on (as far as the floaters do not leave the measurement area) and hence the space averaged concentration $\langle C\rangle$ must remain time independent and proportional to the initial amount of particles poured on the surface, hence so must $\langle \widetilde I\rangle$.
We verified that this property is satisfied for a series of experiments performed at varying mass of glass bubbles poured on the surface with $N=15$ swimmers, for which we computed the spatial average $\langle \widetilde I \rangle(t)$ at each time step. 
Inset of figure \ref{fig:concentration}a shows the resulting spatial average for an initial mass $m=\SI{40}{mg}$ of glass bubbles. 
As expected for $\widetilde I(x, y,t) \propto C(x, y,t)$, the spatial average is indeed time-invariant. The remaining fluctuations have been quantified at large times long after any memory of the initial condition is lost (measurement window in the inset of figure \ref{fig:concentration}a).
The standard deviation are of order $\sigma_{\widetilde I}=10^{-3}$ (this corresponds to relative standard deviation of around 5\,\% for the signal in inset) which serves as an estimate of the error bar of the measurement (smaller than the size of the symbols in figure \ref{fig:concentration}a). 
%

In the next section, we will show that the glass-bubble system reaches a statistically invariant stationary state after an initial stage of few tens of minutes.
In the following, we thus denote by $\overline{Q}$ the time average of a given quantity, with $t\in [55, \,65]\,\mathrm{min}$ corresponding to our measurement window within this late-stage stationary regime.
Figure \ref{fig:concentration}a displays the evolution of the global average concentration $\overline{\langle \widetilde I\rangle(t)}$: it is visible that this quantity evolves linearly with the poured glass bubble mass, validating the assumed proportionality between luminosity and floaters concentration so that equation (\ref{I_propto_C}) can be used to convert recorded images into a concentration field. 

As all moments of the concentration field are expected to scale with the mass concentration, all experiments discussed in the sequel have been performed with the same mass $m=40\pm 2$ mg of glass bubbles which was observed to be a good compromise in having a good signal to noise ratio with no risk of forming multiple layers of bubbles on the surface. 
Note finally that, because the equations of mixing are linear, the results do not depend on the mean concentration, so that in the following we only consider non dimensional concentration fields; 
in order to rub out the very small time fluctuations shown in the inset of figure \ref{fig:concentration}a, the reference chosen is $\overline{\langle C\rangle}$.

\section{Mixing properties}
\label{sec:mixing}

As already mentioned in the previous section, when a large scale patch containing glass bubbles is poured on the surface, it starts to get stretched and folded by the action of the swimmers so that the concentration field $C(x,y,t)$ becomes strongly non homogeneous, as already observed in figure \ref{fig:concentration}b. As the spatial average of the concentration field is conserved, we characterize its heterogeneity by computing the standard deviation 
\begin{equation} 
C_\mathrm{std}(t) = \sqrt{\langle C^2 \rangle (t)-\langle C \rangle^2(t)} \, .
\end{equation}

\begin{figure}
    \begin{centering}
\includegraphics{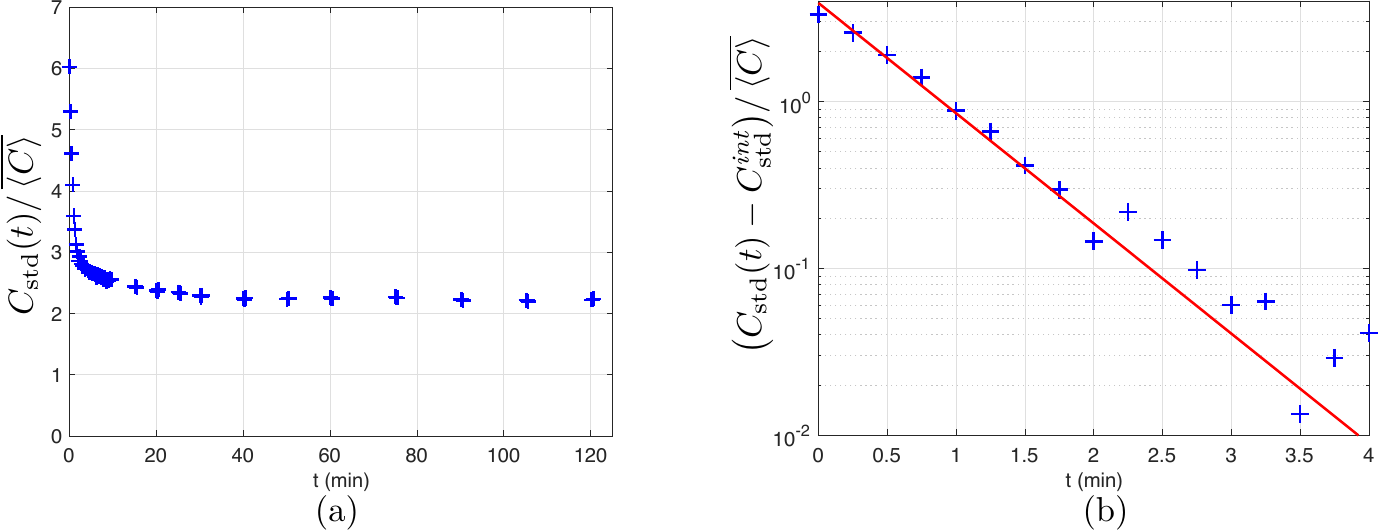}
    \end{centering}
\caption{(a) Temporal evolution of the concentration field  standard deviation: $C_\mathrm{std}(t) = \sqrt{\langle C^2 \rangle-\langle C \rangle^2}$, normalized by the global mean $\overline{\langle C \rangle}$ in the case of $N=15$ swimmers with radius $R=\SI{2.5}{mm}$. (b) Short term evolution of $(C_\mathrm{std}(t)-C^\mathit{int}_\mathrm{std})/\,\overline{\langle C \rangle}$ plotted in semilog representation, with $C
^\mathit{int}_\mathrm{std}/\,\overline{\langle C \rangle}=2.7$. For those figures the points are obtained with a sliding window average over a few seconds, so that the error is very low.}
\label{fig:ectype}
\end{figure}
Figure \ref{fig:ectype}a shows the evolution of $C_\mathrm{std}(t)/\,\overline{\langle C \rangle}$ in the case of $N=15$ swimmers, which is typical of all explored parameters. 
Starting from a finite initial value corresponding to the initial patch, the standard deviation relaxes over a short time scale (less than one minute) as expected for a system being mixed by the random motions of stirrers --- here the swimmers. 
However, while classical expectation would be to decrease down to a stationary fully homogeneous system with $C_\mathrm{std} = 0$, the system reaches a statistically stationary state of incomplete mixing as quantified by $C^\infty_\mathrm{std}$ being more than twice larger than the global mean $\overline{\langle C \rangle}$. 
The mixing process then consists in three main stages, a rapid phase ($t \leq 4$ min) during which the standard deviation $C_\mathrm{std}(t)$ decreases exponentially toward an intermediate value $C^\mathit{int}_\mathrm{std} \simeq 2.7 \overline{\langle C \rangle}$ (figure \ref{fig:ectype}b); 
this exponential decay, typical of chaotic or turbulent mixing \cite{bib:boylandetal2000,bib:Gouillartetal2006,bib:tennekes_Lumley1972}, is followed by a slower relaxation ($t \in [4-40]$ min) and finally a long phase ($t \geq 40$ min) for which the statistical properties of the concentration field only weakly evolve due to a slow loss of activity of the swimmers.
All characterizations (including calibration as discussed in the previous section) have been performed in this third phase using a $10$ min recording to ensure that the swimmers activity remains the same when comparing experiments, with stationary statistical properties.

\begin{figure}
\includegraphics{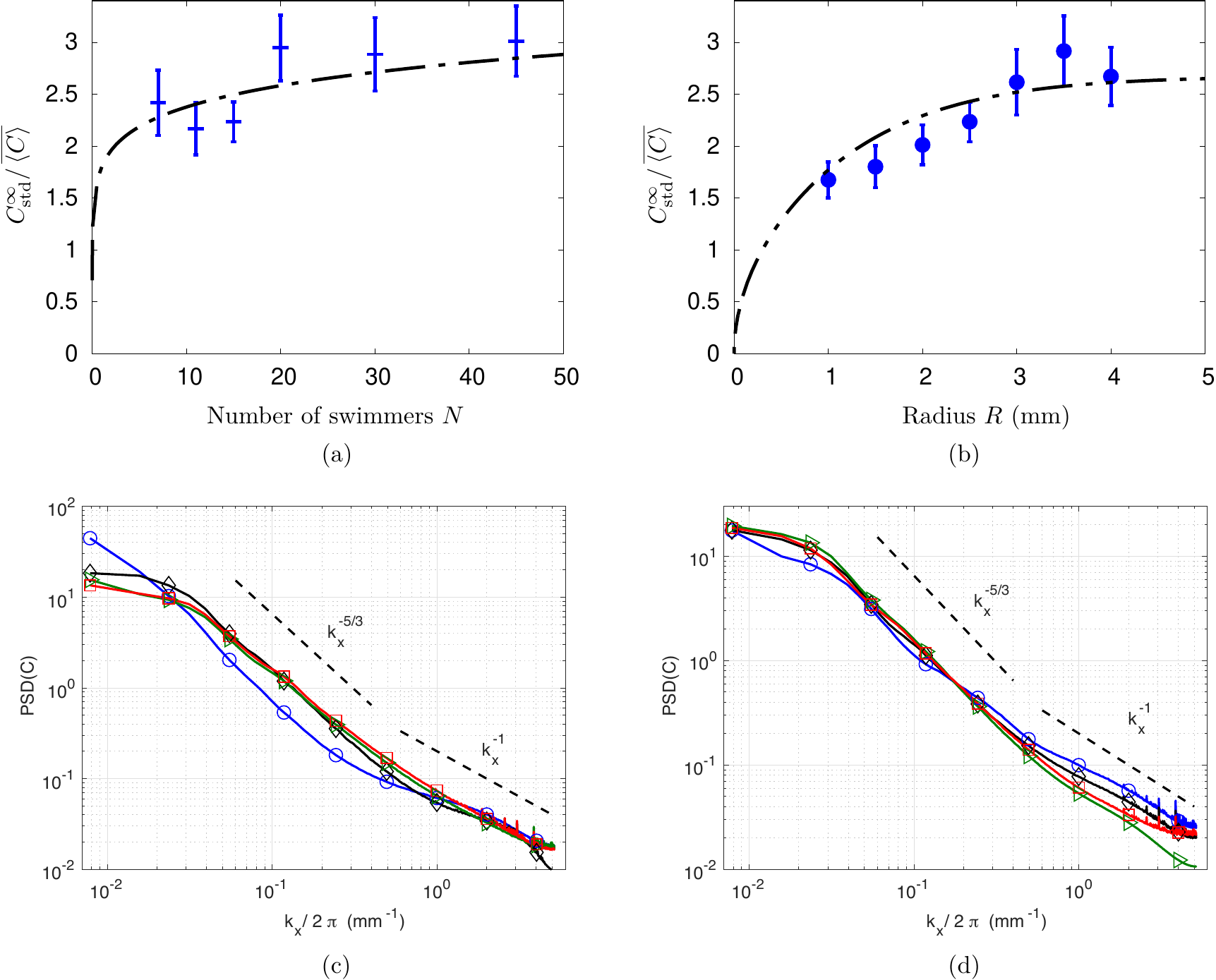}
      \caption{\textit{Top:} Evolution of the standard deviation of the concentration in glass bubbles $C_\mathrm{std}^\infty/\overline{\langle C\rangle}$ as a function of (a) the number of swimmers $N \in [7,\,45]$ with radius $R=2.5$ mm; (b) the radius $R \in [1,\,4]$ with $N=15$ swimmers.  In both figures, the fit (dashed dotted line) results from a model explained in section \ref{subsec:model_Cstd}, and the error bars were calculated using a logarithmic differentiation.     
      (c) 
      Power spectrum density of the glass bubble concentration as a function of the wavenumber $k_x$ for increasing number of swimmers with $R=2.5$ mm. $(\circ)$ $N=7$ swimmers, $(\diamond)$ $N=15$ swimmers, $(\triangleright)$ $N=30$ swimmers, $(\square)$ $N=45$ swimmers. Dashed lines correspond to power law spectra with exponent $-5/3$ and $-1$. 
      (d) Power spectrum density of the glass bubble concentration as a function of the wavenumber $k_x$ for $N=15$ swimmers and different radii $R$. $(\circ)$ $R=1$ mm, $(\diamond)$ $R=3$ mm, $(\triangleright)$ $R=3$ mm, $(\square)$ $R=4$ mm. Dashed lines correspond to power law spectra with exponent $-5/3$ and $-1$.}
      \label{fig:spectres}
\end{figure}

Surprisingly, when investigating how this final mixing degree $C^\infty_\mathrm{std}$ depends on the mixer properties (number and size of camphor swimmers), it is found to be quite robust and independent on the conditions. Indeed, Fig. \ref{fig:spectres} (a,b) report that except for the smallest particle sizes ($R<\SI{2.5}{mm}$) or numbers ($N<15$) for which the final state is slightly closer to homogeneous, the final standard deviation of concentration remains unaffected.

%
It should be noted that the standard deviation observable yields a global characterization that integrates over concentration fluctuations at all length scales.
A simple look at figures \ref{fig:setup}b or \ref{fig:concentration}b however unveils a rich underlying spatial organization of the floaters with (i) large void features around each swimmer typical of the present system together with (ii) thin and complicated structures in between voids  qualitatively reminiscent of classical mixing processes.
Indeed, in a recent study the dynamics of the camphor swimmers was shown to exhibit multi-scale features following typical turbulent scaling laws~\cite{bib:bourgoin_etal2020}. 
This contribution caused by the swimmers relative velocities is responsible for the floaters mixing that shows up in between void features. We now explore the multi-scale nature of this mixing process and shall come back in the next section on the mechanism responsible for the void generation around each swimmer.

%
%
%
%
%

To characterize the multi-scale mixing, we compute one dimensional power spectra of the concentration field along the $x$ axis
\begin{equation}
|\hat{C}|^2(k_x,y,t)=|\mathrm{DFT}_x[C(x,y,t)]|^2,
\end{equation}
where DFT stands for Discrete Fourier Transform.
It is computed in a square box of size 1310x1310 pix$^2$ centered in the middle of the surface, and is a function of the wave number $k_x$, $y$, and $t$, so that we average it over space and time in the stationary regime to get a better statistics. 
Figures \ref{fig:spectres} (c,d) display the corresponding power spectra, $\overline{\langle |\mathrm{DFT}_x[C(x,y,t)]|^2\rangle_{y}}(k_x)$, obtained when increasing the number of swimmers at fixed radius $R=\SI{2.5}{mm}$, or for different radii at fixed $N=15$. 
It is visible in these two figures that the concentration field exhibits fluctuations at all spatial scales whatever the number of swimmers or their radii. 
When increasing the number of swimmers at fixed $R$, the power spectra are attenuated in the low wave numbers range $k_x \leq 0.05$ mm$^{-1}$ while exhibiting higher and higher fluctuations in the high wave number range $k_x \geq 0.05$ mm$^{-1}$. 
This is inline with the fact that the global mixing efficiency does not vary much when increasing $N$  as $(C_\mathrm{std}^\infty)^2$ is proportional to the area under each curve. 
When $N$ is large enough so that the turbulent-like behavior of the swimmers develops \cite{bib:bourgoin_etal2020}, the concentration field is efficiently stretched and folded, which results in small spatial structures. 
In that regime ($N \geq 15$), we observe that concentration spectra tend to follow a power law behavior with an exponent close $-5/3$ in the intermediate spatial frequency range as observed in hydrodynamic turbulence \cite{bib:Warhaft2000}. 
It is remarkable that a second regime emerges in the very high frequency range where scalar spectra exhibit a second power law behavior with an exponent close to $-1$. 
This Batchelor type spectra \cite{bib:batchelor1959} indicates that the flow is smooth at these spatial scales so that small scales are created via random advection as observed in experiments and numerical simulations  \cite{bib:Williansetal1997,bib:pierrehumbert1994,bib:toussaintetal2000}. 
Figure \ref{fig:spectres}\,(d) shows that all the aforementioned results appear to be modulated by the radius of the swimmers: (i) increasing the radius $R$ of the particles, which results in an increase of the swimmer Reynolds numbers, reinforces the power law behavior at intermediate scale, which extends over more than one decade when $R=4$ mm; (ii) the exponent in the high frequency range gets closer to $-1$ when increasing the radii of the swimmers although such regime was not very well developed with $R=2.5$ mm.

Because the system reaches an out-of-equilibrium steady state, a competing mechanism must balance with the aforementioned mixing processes. 
Indeed, as already pointed each swimmer trails an area devoid of floaters, suggesting the existence of an \textit{unmixing} mechanism that constantly rejuvenates large-scale heterogeneities. 
Those empty wakes eventually feed the mixing process to smaller scales, leading to the non-trivial spectra shown in fig.\ref{fig:spectres}.
The evolution of the wake of individual swimmers is therefore a key ingredient to understand the overall process that we now investigate.

\section{Around a swimmer}
\label{sec:around_swimmer}
\subsection{Averaged concentration field}

As already pointed out, a striking feature of the present system is the wake of washed up surface free from floaters, that follows each swimmer (see figures \ref{fig:setup}b or \ref{fig:concentration}b). 
To get a more quantitative insight into this phenomenon, we now define and consider the mean concentration field around a single swimmer.
To this aim, we perform a coherent mean whereby we average the concentration field in the neighborhood of a single swimmer after a set of geometrical transformations to ensure spatial registration (translation of the swimmer's position) and orientational registration (rotation to provide identical swimming direction). 
In addition, in order to take into account only isolated camphor disks we discard those whose center lies within \SI{2.5}{\centi\metre} from the cell edges, and those from which at least one other swimmer lies in an exclusion rectangular zone.
The size of the exclusion zone (\SI{9}{\milli\metre} in front of a swimmer, \SI{32}{\milli\metre} behind it and \SI{14}{\milli\metre} in directions perpendicular to its trajectory) was chosen according to typical extent of depleted zone as seen in figure \ref{fig:setup}b and \ref{fig:concentration}b. 
Note that extending it further did not change significantly the outcome except for a drastic reduction of the statistics.
Overall, a typical set of 10\,000 images of swimmers' neighborhood was used for computing the averaged concentration field of floaters.
\begin{figure}
    \centering
\includegraphics{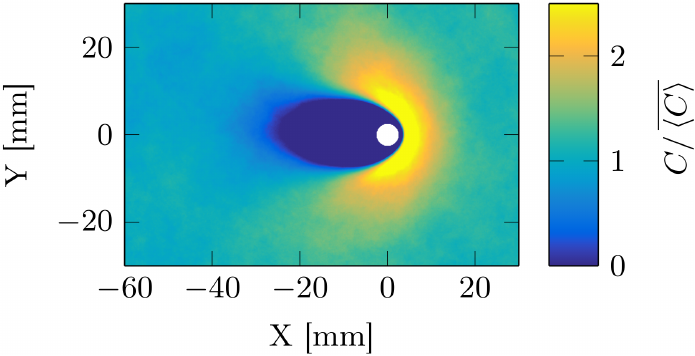}
    \caption{Averaged glass bubbles concentration field around a swimmer (in white), going from the left to the right, obtained by a coherent mean over 10\,000 concentration fields. 
     The configuration considered here is $N=15$ and $R=$\SI{2.5}{\milli\metre}. 
    \label{fig:immoy}
    }
\end{figure}
Figure \ref{fig:immoy} shows such an averaged field around a swimmer, going from left to right on the figure, here in a configuration with $N=15$ camphor swimmers of radius $R=\SI{2.5}{mm}$. 
Two remarkable features are noticeable: (i) as expected from figure \ref{fig:concentration}b, a depleted wake trailing behind the swimmer where all floaters have been swept away, and (ii) an accumulation front immediately ahead of the swimmer, that was far less visible. 
This type of pattern is typical of all conditions, see figures \ref{fig:meanTrailvarN} and \ref{fig:meanTrailvarR} in the appendix, for different numbers of swimmers $N$ and radii $R$.

\begin{figure}
    \centering
\includegraphics{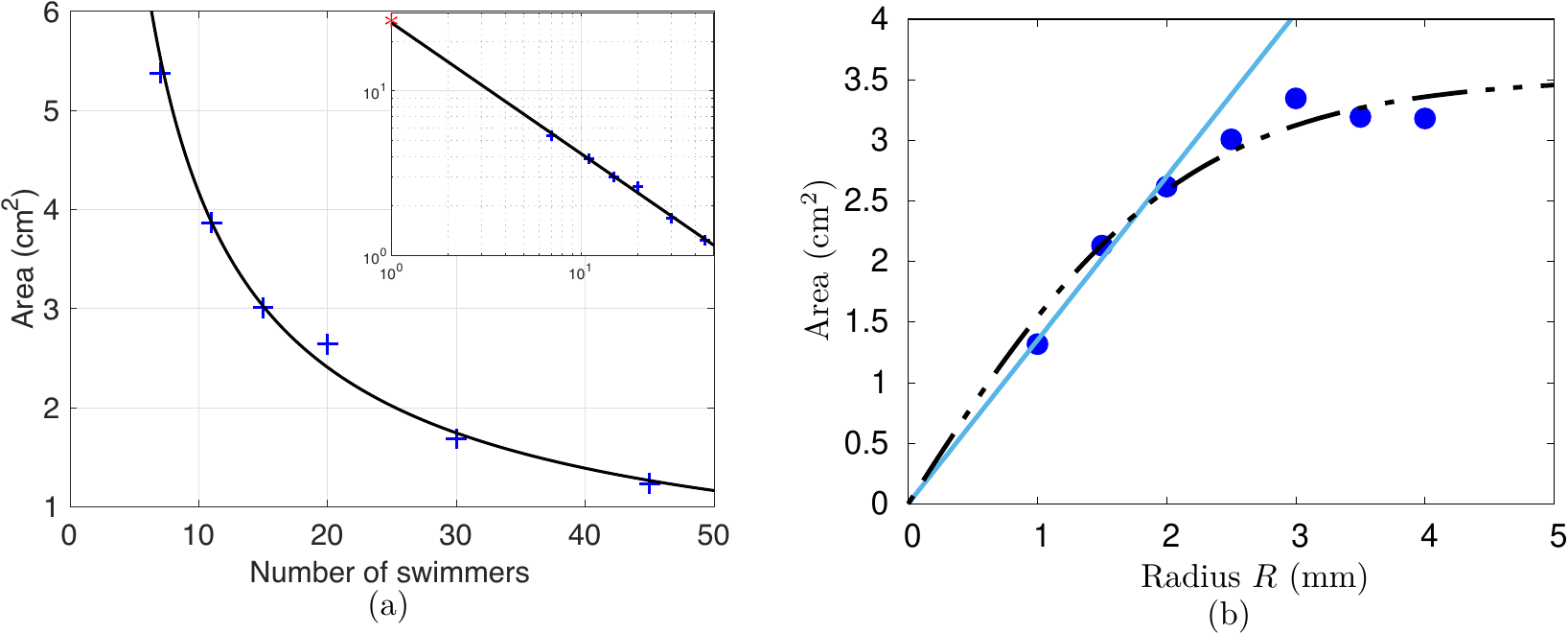}
      \caption{
            (a) \textcolor{blue}{\textbf{$+$}}: Measured depletion area $A_d$ as a function of the number of swimmers $N$ (fixed radius R=\SI{2.5}{mm}). Inset: same in log-log scale; the point \textcolor{red}{$\ast$} for $N=1$ is measured in different conditions, see later in figure \ref{fig:TestOrigin}b. Solid line: fit by a power-law decay $A_d(N,R=\SI{2.5}{mm})=A_d^{(1)} \, N^{-0.76}$, with $A_d^{(1)} = \SI{24}{\centi\meter\squared}$.
            (b) \textcolor{blue}{$\bullet$}: Measured depleted area $A_d$ as a function of the radius $R$ of the swimmer, for $N=15$ swimmers in the bath. Solid line: linear behavior at small radii; dashed line: guide-line fit by hyperbolic tangent function yielding $A_d=3.52\,\tanh{(0.47R)}$ in \si{\centi\meter\squared}.
}
      \label{fig:deplarea}
 \end{figure}

From the averaged concentration fields obtained for all configurations (Fig. \ref{fig:immoy} and appendix figures \ref{fig:meanTrailvarN} and \ref{fig:meanTrailvarR}), we extract the total wake area by fitting with an ellipsoidal shape.
Note that additional information on the depleted surface can be found in appendix \ref{appC}.
This measured depleted area $A_d$ is shown in figure \ref{fig:deplarea} as a function of the number of swimmers $N$ (Fig. \ref{fig:deplarea} a with $R=\SI{2.5}{mm}$), and as a function of the radius $R$ of the swimmer (Fig. \ref{fig:deplarea}b with $N=15$ swimmers in the bath).
As clearly seen in figure \ref{fig:deplarea}a, the depleted area around each swimmer decreases with the number $N$ of swimmers. 
This decrease exhibits a well defined power-law behavior (see figure inset for the log-log scale) that yields a fitted exponent close to $-3/4$, \textit{i.e.}  $A_d(N,R=\SI{2.5}{mm})=A_d^{(1)} \, N^{-0.76}$,
with $A_d^{(1)} = \SI{24}{\centi\meter\squared}$ the fitted wake area of a single isolated swimmer.
On the contrary, the depleted area is found to increase with the radius of the swimmers $R$ at fixed density ($N=15$).
In practice, the increase might be viewed as close to linear at the smallest sizes, before to saturate at larger radii. For practical purposes, a fit by an hyperbolic tangent function reasonably approximates the overall behavior (figure \ref{fig:deplarea}b).

\subsection{A first order model for the standard deviation of concentration}
\label{subsec:model_Cstd}
Besides the excluded area, another striking feature from figure \ref{fig:immoy} is the fact that the concentration relaxes rapidly towards $\overline{\langle C\rangle}$ when moving away from the camphor swimmer. 
This is another evidence that mixing by the moving particles is efficient, inline with the spectra shown in figure \ref{fig:spectres}. 
Therefore one can wonder whether at first order, the large-scale inhomogeneity measured through $C_\mathrm{std}$, of the order of $1.5$--$ 3\,\overline{\langle C\rangle}$ in figure \ref{fig:spectres}, could primarily reflect the patchiness of the superposition of the wakes and accumulation fronts of each swimmer, and therefore be related to their respective areas. 
According to this proposition a simple estimate for $C_{std}$ can be developed as follows. 

Let us consider $N$ identical swimmers of radius $R$ carrying each a depleted wake of area $A_d$, an accumulation front ahead of the swimmer of area $\alpha\,A_d$, and let $A_T$ be the total area of the system. 
As before, we neglect the area of the swimmers. 
Let us assume for the sake of simplicity 
that the wakes and accumulation fronts have identical respective extents for each swimmer and that they do not overlap.
We therefore model the surface and concentration distributions as follows:
\begin{itemize}[label=$-$,leftmargin=1cm ,parsep=0cm,itemsep=0cm,topsep=0cm]
\item $N$ depleted area of total surface $NA_d$, having a zero concentration; 
\item $N$ accumulation fronts ahead of the swimmers of total area $N\alpha A_d$, in which the over concentration comes from glass bubbles that are not in the wakes, that is a concentration $(1+\alpha)A_d/(\alpha A_d)\langle C\rangle$;  
\item a uniform concentration equal to $\langle C\rangle$ everywhere else, on an area equal to $A_t-N(1+\alpha) A_d$. 
\end{itemize}
We have:
\begin{eqnarray}
\left(C_\mathit{std}^\infty\right)^2&=&  \langle C^2\rangle-\langle C\rangle^2\\
      &=&  \frac{1}{A_t}\left[0+N\alpha A_d\left(\frac{1+\alpha}{\alpha}\right)^2\,\langle C\rangle^2+\bigl(A_t-N(1+\alpha)A_d\bigr) \langle C\rangle^2 \right]-\langle C\rangle^2\\
&=&  \frac{N A_d}{A_t}\left[\frac{(1+\alpha)^2}{\alpha}-(1+\alpha) \right]\langle C\rangle^2\\
&=&  \frac{N A_d}{A_t}\, \frac{1+\alpha}{\alpha}\, \langle C\rangle^2
\end{eqnarray}
and finally
\begin{equation}
C_\mathit{std}^\infty=\sqrt{\frac{N A_d}{A_t}\, \frac{1+\alpha}{\alpha}}\, \langle C\rangle\,.
\label{eq:Cstd}
\end{equation}

We now confront this first order model against previous experimental measurements. 
To do so, let us note that measured standard deviations are obtained from instantaneous images, and then averaged over time. 
Because the tails of depleted wakes are often curved in a random direction, the procedure for calculating the averaged concentration field around a swimmer thus rubs out wakes and eventually leads to an underestimation of their extent.  
Compared with instantaneous images, this underestimation amounts to about 50\% depending on the swimmer considered and the given time.
This requires a correcting factor in Eq.~(\ref{eq:Cstd}) for direct comparison with data. 
Alike, the concentration averaging smooths the maximum concentration ahead of the swimmer: from about $6\,\langle C\rangle$ on raw images depending on the instant $t$ chosen (Fig. \ref{fig:concentration}b), down to $2.5\,\langle C\rangle$ (Fig. \ref{fig:immoy}) in averaged concentration fields. 
Accordingly, the parameter $\alpha$ setting the size of the accumulated area in the model is chosen to be $\alpha=0.2$ to yield the proper maximum concentration value ($6 \langle C\rangle$). Such an extent of 20\,\% of the depleted area is consistent with direct observations (Fig. \ref{fig:concentration}b).

Qualitatively, using the $N$-dependency of the averaged concentration field found in Fig.~\ref{fig:deplarea}, our simple model Eq.~\ref{eq:Cstd} predicts a measured standard deviation for the glass bubbles behaving as $C_\mathrm{std}\propto N^{0.12}$. 
This very weak power-law prediction is consistent with the very slow evolution reported in Fig. \ref{fig:spectres}a, for $N\ge7$.
As for the $R$-dependency, the initially linear evolution of the single swimmer averaged depleted area (Fig. \ref{fig:deplarea}b) followed by a saturation should lead to $C_\mathrm{std}\propto\sqrt{R}$ before reaching a plateau.
Again, this is fairly consistent with the observed trend in figure \ref{fig:spectres}b.

In a more quantitative way, we can fit the measured data figures \ref{fig:spectres}a,b by our model expression including the previously mentioned correcting factors
\begin{equation}
  C_\mathit{std}^\infty/\overline{\langle C\rangle}=\beta\,\sqrt{\frac{N \,A_d^\mathrm{eff.}}{A_t}\, \frac{1+\alpha}{\alpha}}\,,
\label{eq:Cstd_fit}
\end{equation}
with $\beta$ a free scaling parameter, $\alpha=0.2$ and $A_d^\mathrm{eff.} = 1.5 A_d$, where $A_d(N, R)$ is the averaged depleted area around a single swimmer. 
We used this expression, combined with the equations given in figure \ref{fig:deplarea} for the behavior of $A_d$ versus $N$ or $R$, in order to plot 
the fits shown in dotted lines in figures \ref{fig:spectres}a and b.
As can be seen, it captures in both cases the overall experimental behavior with a scaling factor $\beta=1.95$.
This is a very reasonable order of magnitude for such a simple model, where only three regions (each with a given value of concentration) are considered.
Because the evolution of $C_\mathrm{std}$ with $R$ and $N$ reflects that of the wakes of individual swimmers, $C_\mathrm{std}$ is phenomenologically dominated by demixing at large scale; it is thus reasonable to associate the demixing properties of this flow to the strongly inhomogeneous concentration field around the swimmers.


\section{Discussion: Marangoni effects}
\label{sec:Marangoni}

Up to now we have mainly focused on the mixing/demixing properties of the flow, associated with the depleted areas around the swimmers, regardless of the associated physical process that produces them. 
We now discuss the different possible origin for this phenomenon, and show how the description of Marangoni effects accounts consistently for the observed behaviors.  

\subsection{Origin of the depleted area}
\label{sec:origin_depleted}
Physically, it is tempting to link the existence of the depleted area to the chemical cloud released by a single swimmer and to the associated Marangoni flows driving the self-propulsion. This, however, requires that we discard other plausible origins among which the possible wake around a disk moving at finite Reynolds number along the interface. 
\begin{figure}
    \centering
\includegraphics{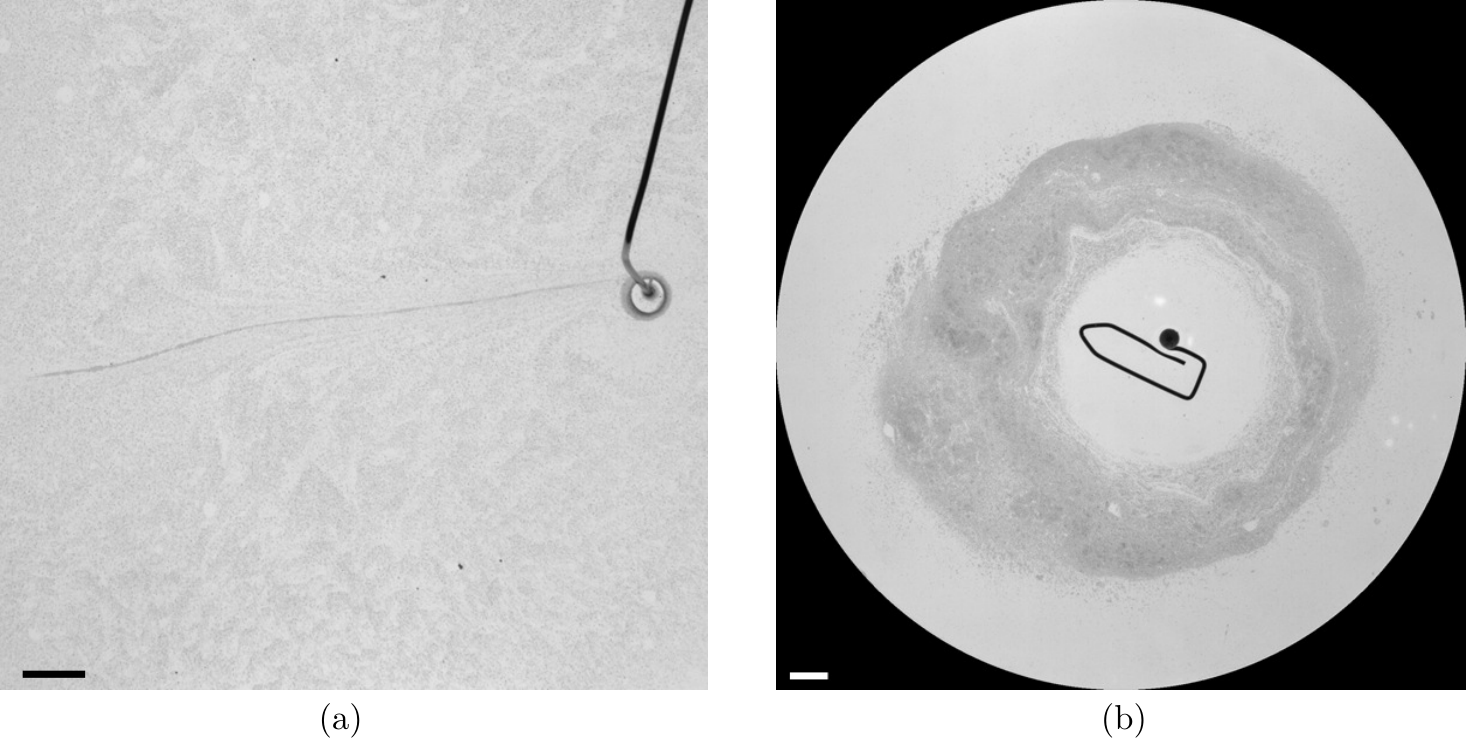}
      \caption{(a) An agarose disk of radius \SI{2.5}{mm} is pulled at \SI{6}{\centi\metre\per\second} by an engine. 
      No depleted area is observed, proving that mechanical effects can not explain what we observe in Fig. \ref{fig:immoy}.
      Scale bar is \SI{1}{cm}.
      (b) An interfacial swimmer is fixed in the bath. Glass bubbles are poured on it.
      In less than a second, a large depleted area is observed, leading us to think of a chemical effect.
      Scale bar is \SI{1}{cm}.}
      \label{fig:TestOrigin}
\end{figure}

To discriminate between these scenarios, we performed two complementary experiments:  
in a first configuration, an agarose gel disk \textit{without camphor loading} is moved along the air-water interface at a constant velocity imposed by a motorized translation stage. 
At a velocity typical of the swimmers velocity $U = \SI{6}{\centi\metre\per\second}$, the simple motion of the disk at an interface filled with glass bubbles floaters does not generate any significant pattern around the moving disk (Fig. \ref{fig:TestOrigin}a).
In particular, no signature of a swept wake is observed, the only feature being a thin concentrated filament released at the rear, due to floaters trapped at the disk edge by capillary effects. 
Note also that no accumulation front is visible ahead of the moving disk in this configuration.
In the second configuration, we cancel the disk motion by nailing it at a fixed position at the interface but restore the camphor loading of the disk so that chemical release and associated Marangoni flows do occur: a clear pattern develops around the disk where all floaters in its vicinity are swept away and leave a depleted area reminiscent of the one observed with swimmers. 
Such an observation is consistent with closely related experiments with Marangoni driven flows by surfactant spreading \cite{bib:roche_etal_PRL2014,bib:leRouxetal2016} where a similar cleaned-up surface-area is observed. 
The area obtained here (around \si{26\,\centi\meter^2}), although of larger extent than in Fig. \ref{fig:immoy}, is in very good agreement with the value given by the fit in figure \ref{fig:deplarea}a for $N=1$ (\si{24\,\centi\meter^2}).

Overall, the camphor spreading from each disk triggers two different effects which act oppositely on the mixing of floaters: first, it generates the net motion of each disk, all the disks operating as an assembly of stirrers moving randomly in the system. Second, the complex Marangoni flow pattern attached to each swimmer induces a local unmixing mechanism that constantly rejuvenates gradients by sweeping of floaters from a finite area, so as to form a concentrated rim ahead of the swimmer, together with an empty wake.

In the next two subsections, we examine how the description of Marangoni effects can indeed rationalize some of the observations reported earlier.

\subsection{Single Swimmer}
We first discuss how the generation of the depletion area around each swimmer can be captured based on Marangoni flow description.
As a first step, we begin with an estimate of the camphor concentration field generated by a single swimmer of radius $R$ and typical velocity $U$. 
The viscous friction is $F_v = \pi R^2 \eta U / \delta_v$ \cite{bib:ockendon1995}, where 
\begin{equation}
  \delta_v = R/\sqrt{\mathrm{Re}} 
  \label{eq:delta_v}
\end{equation} 
is the viscous boundary layer thickness underneath the swimmer, with $\mathrm{Re}=UR/\nu$ the Reynolds number.
Balancing $F_v$ with the driving capillary force contribution $F_c = \pi R \Delta\gamma$, where $\Delta\gamma$ is the fore-aft surface tension difference, we obtain
\begin{equation}
  \Delta\gamma \sim \eta\, U \sqrt{\mathrm{Re}}\,.  
\label{eq:dGamma}
\end{equation}
For our typical situation with $R=\SI{2.5}{\milli\meter}$ and
$U\approx\SI{6}{\centi\meter\per\second}$ for $N=1$ swimmer \cite{boniface2019self}, corresponding to a Reynolds number $\mathrm{Re}=150$, 
this leads to a typical surface tension imbalance of $\Delta\gamma\simeq\SI{0.7}{\milli\newton\per\meter}$ in good agreement with previous experimental estimates on camphor boats \cite{Karasawa:2014gq}.
To proceed we now assume, in agreement with the literature, a linear relationship $\Delta\gamma = - \alpha C$ between surface tension and local camphor concentration, with $\alpha = \SI{3e-3}{\newton\square\meter\per\mol}$ \cite{Soh_jpcb-2008}. 
We obtain a characteristic concentration behind the swimmer $C^*\simeq \SI{0.2}{\mol\per\cubic\meter}$, far below the solubility limit  $C_\mathrm{sat.} = \SI{8}{\mol\per\cubic\meter}$.
Qualitatively, this camphor release in the vicinity of the swimmer induces, on top of the capillary force, a Marangoni stress at the free surface, from low to high surface tension region, that drives fluid outward, away from the swimmer. 
This flow sweeps away surface floaters, generating a clean depleted area around each swimmer (see Fig. \ref{fig:TestOrigin}b). 
To make this argument more quantitative, it is natural to identify the depleted area to the camphor-contaminated area over which Marangoni stress occurs. 
We also simplify the problem by considering a fixed camphor particle (as in figure \ref{fig:TestOrigin}b), yielding to a depleted disc (rather than an ellipse) of area $A_d$ and radius $R_d$.
In that case the typical velocity of Marangoni flows produced matches with the freely moving swimmers velocity $U$ in agreement with experimental measurements \cite{Sur:2019bj}; 
therefore, except for the simplified geometry, all physical scalings remain identical.

On the one hand the swimmer releases camphor at a rate proportional to its area; this production term writes $Q_p = \beta R^2$,  with $\beta =\SI{3.1e-4}{\mol\per\second\per\square\meter}$ as measured for this system \cite{boniface2019self}. 
On the other hand camphor is removed from the surface by dissolution \footnote{Note that for camphor, the evaporation/sublimation flux towards the upper atmosphere can act as a competing removal mechanism. Such an alternative route is discussed in appendix \ref{ap_sec:camphor} and does not change the overall picture described here.}, at a rate 
\begin{equation}
    Q_d = D \frac{C^*}{\delta_D} A_d,
\label{eq:CamphorConserv}
\end{equation}
with $\delta_D$ the thickness of the diffusion boundary layer $\delta_D = (R_d/R)^{1/2}\;\delta_v / \sqrt{\mathrm{Sc}}$ \cite{bib:levich1962}, $\mathrm{Sc} = \nu/D$ the Schmidt number, and $D=\SI{7e-10}{\square\meter\per\second}$ the camphor diffusivity \cite{boniface2019self}. 
In line with recent treatments of other Marangoni spreading problems \cite{Mandre_jfm-2017, bib:roche_etal_PRL2014}, we suppose that production is balanced by dissolution according to
\begin{equation}
    Q_p = Q_d\,.
\label{eq:produc_dissip}
\end{equation}
This yields to an extension of the depleted area
\begin{equation}
    R_d = 
    \left[
        \frac{\alpha\beta}{\pi\rho D \;\mathrm{Sc}^{1/2}}
    \right]^{2/3}
    \;\frac{R}{U^{4/3}}.
\label{eq:Ldiff_iso}
\end{equation}
With the typical values given above, this predicts a contaminated area of extent $R_d\simeq 3R$, to be compared with the measured extent $R_d\simeq10R$ (Fig~\ref{fig:TestOrigin}b). Considering the rough scaling approach used, this shows a very fair agreement although it underestimates the size of the depleted area. 

Finally, noticing that the swimming velocity of individual swimmers was shown to follow a scaling law of the form $U\propto R^{1/3}$ \cite{boniface2019self}, it is possible to gather all radius-dependencies in Eq. \eqref{eq:Ldiff_iso} to reach a theoretical expectation that $R_d\propto R^{5/9}$, that is, a depleted area $A_d\propto R^{10/9}$. 
While this result has been derived for a --fixed-- isolated swimmer, experimental measurements always involve multiple swimmers. 
For a single swimmer, it is indeed not possible to properly define a wake as the camphor disk gets trapped at tank edges where it moves along the outer perimeter.
This implies that measured wakes for multiple swimmers are expected to follow the above scaling only in the limit of small interaction and overlap among wakes.
For a fixed number of swimmers $N=15$, this is best achieved for small wakes corresponding to small size swimmers. 
Indeed figure \ref{fig:deplarea}b shows an almost linear trend for small $R$, consistent with the previous scaling law, before the depleted area eventually saturates at larger radii for which wakes overlap.

Overall, simple estimates considering Marangoni effects can successfully account for the dependence of the depletion area with the radius of the swimmer and predict with a rather good order of magnitude its size.

\subsection{Multi Swimmers}
While previous arguments were developed for isolated swimmers, we now focus on the wake behavior with multiple swimmers.
In crowded environment when the mean distance $d$ between two swimmers, scaling as $d \sim \sqrt{A_t/N}$, starts to compare with the isolated wake extension, one naturally expects that the typical radius of the depleted area should be limited by $d$ and thus decrease with increasing $N$.
The inset in figure \ref{fig:deplarea}a however suggests that the wake area is a \textit{strictly} decaying function of $N$, even for small values of $N$ for which this naive crowding effect should not contribute. 
Indeed, in the case of 7 swimmers of radius $R=\SI{2.5}{\milli\meter}$, the area of one wake amounts to \SI{5.4}{cm^2}. 
This is significantly smaller than the anticipated isolated swimmer wake (\SI{26}{cm^2} for $N=1$) despite consisting on a hardly crowded configuration: the total accessible area  $A_t = \SI{254}{cm^2}$ should allow to host 7 swimmers with wakes area equal to \SI{26}{cm^2} as for isolated swimmers.

In order to test whether this decay could be attributed to Marangoni effects, we propose a very simple 1D analytic model that mimics this situation. 
Let us consider a 1D system of fixed equidistant chemical sources that correspond to our swimmers. For a finite size system of width $W$, the inter-source distance $d$ thus goes like $d\approx W/N$. In the following, we will neglect edge effects by considering an infinite system, keeping in mind that $1/d$ reads for the number $N$ of swimmers.

We now assume that each single source --\textit{i.e.} camphor particle-- generates a camphor distribution around its location $x_0$ that is an even function of $x-x_0$, with standard deviation $\sigma$, and decreasing with distance to the origin $x_0$. This is due to camphor being spread and eventually lost from the surface by several effects (Marangoni effect, dissolution, sublimation, diffusion)
The precise shape is not crucial; for simplicity we choose a  Gaussian profile $C(x)=C_0 \exp(-x^2/2 \sigma^2)$. 
We further assume that when placing camphor particles on a line, the total camphor concentration $C_t$ is a linear superposition of the effects from all sources. 
One then has:
\begin{equation}
C_t(x) = \sum_{n\in \mathbb{Z}} C(x-n d)\,.
\end{equation}
Once the camphor concentration is known, we take care of the distribution of glass bubbles on the surface, and denote by $G_b(x)$ the corresponding concentration. 
Glass bubbles are repulsed by the camphor due to the Marangoni flow through a compressible velocity of the type 
\begin{equation}
     v = - \alpha \partial_x C_t.
\label{eq:v_alpha_partial_C}
\end{equation}
This repulsive contribution is balanced by a diffusive transport term with coefficient $D_b$ that tends to homogenize glass bubbles so that the conservation equation for the distribution $G_b$ satisfies:
\begin{equation}
\partial_x \bigl(v\, G_b - D_b\, \partial_x G_b\bigr) = 0\, ,
\end{equation}
with $v$ solution of equation (\ref{eq:v_alpha_partial_C}). 
One then gets 
\begin{equation}
v(x)\, G_b(x) - D_b\, \partial_x G_b(x) = B\, .
\label{eq:eq_with_B}
\end{equation}
The value of the constant can be obtained by a symmetry argument: 
indeed, the solution is periodical with period $d$ so that one can take a local average of the previous equation.
Defining the average as
\begin{equation}
\langle f \rangle_d(x) = \frac{1}{d} \int_{x-d/2}^{x+d/2} f(x') dx'\, ,
\end{equation}
one gets $\langle v\, G_b \rangle_d = B$. 
In the case a single swimmer ($d\longrightarrow+\infty$), $C(x)$ is an even function, 
so that for reasons of symmetry, $G_b(x)$ is also an even function;
because $v$ is an odd function (equation \ref{eq:v_alpha_partial_C}),  
 one gets $\langle v\, G_b \rangle_d (x=0)= 0 = B$. 
Note that this corresponds to a solution with zero mean flux of particles transported by the flow. Actually, this term has to vanish whatever the symmetries of $C(x)$ if glass bubble are confined in a box as they can not leave the domain, so that the total flux vanishes at the boundaries.

Equation \ref{eq:eq_with_B} now writes:
\begin{equation}
D_b\, \partial_x \log G_b(x) = -\alpha \partial_x C_t(x) 
\end{equation}
with general solution
\begin{equation}
G_b(x) = G_0 \exp(-\alpha C_t(x)/D_b),
\label{eq:G_b}
\end{equation}
a result that was checked numerically using Monte-Carlo simulations.

\begin{figure}
\includegraphics{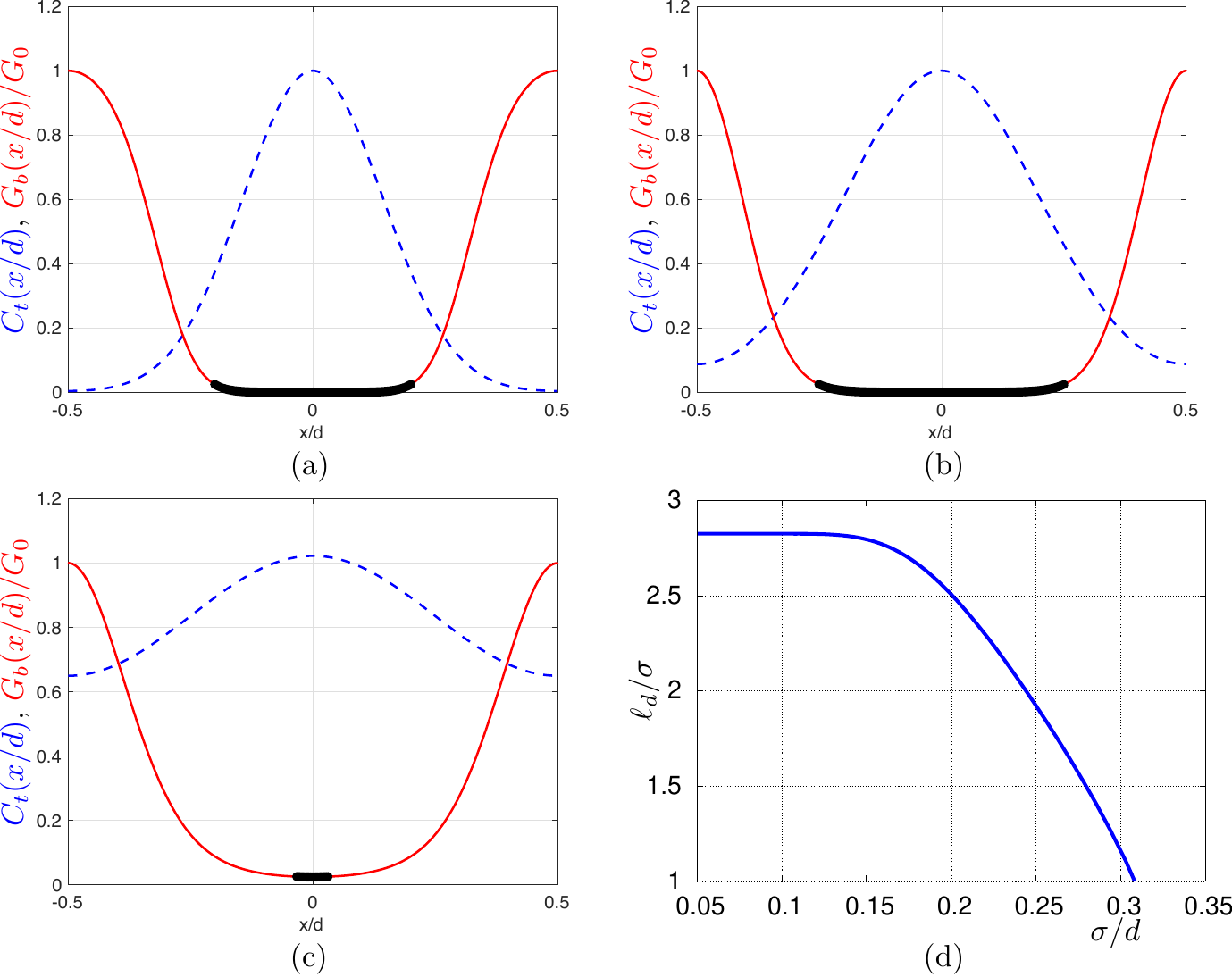}
   \caption{a, b, c : Camphor concentration $C_t(x)$ (dashed line), glass bubbles concentration $G_b(x)/\max(G_b)$ (solid line) as a function of $x/d$ for $d/\sigma=7$, $5$, $3$. $G_b(x)/\max(G_b)$ is given by Eq. \ref{eq:G_b} with $\alpha C_0/D_b=10$. The depleted zone, represented by black symbols, corresponds to $G_b < 0.025 \max(G_b)$. d: Evolution of the dimensionless depletion length $\ell/\sigma$ as a function of $\sigma/d\propto N$.}
    \label{fig:1Dmodel_Marangoni}
\end{figure}
Figures \ref{fig:1Dmodel_Marangoni}a, b, c display $C_t$ and $G_b/\max(G_b)$ as a function of $x/d$ for a $d$-periodic distribution of Gaussian profiles and different ratios of $d/\sigma$.  
On those profiles the ratio $\alpha C_0/D_b$ is set to 10, and the size of the depleted zone $\ell_d$ is defined  as the region where the glass bubble concentration, $G_b(x)$, is smaller than a threshold set to $2.5\%$ of its maximal value 
\footnote{Because of the exponential in equation \ref{eq:G_b}, changing the threshold does not change much the size of the depleted region provided that it is not too small.}.
The depleted region corresponds to the black symbols in figures \ref{fig:1Dmodel_Marangoni}a, b, c, and is represented in figure \ref{fig:1Dmodel_Marangoni}d as a function of $\sigma/d$ for $3\le d/\sigma\le 20$.

For large enough values of $d$, the camphor clouds of two neighbouring particles should not interact much;
this is illustrated in figure \ref{fig:1Dmodel_Marangoni}a for $d=7\sigma$ (corresponding to $\sigma/d\simeq 0.14$), where the total concentration of camphor nearly goes to zero at $x=d/2$. 
For larger values of $d$ we therefore expect the size $\ell_d$ of the depletion region to be essentially unchanged: this is exactly what is observed in figure \ref{fig:1Dmodel_Marangoni}d with the plateau for $\sigma/d< 0.14$. 
For smaller gaps between swimmers, the wakes interact, the camphor distribution is less steep with an increasing minimum of $C_t$ (figures \ref{fig:1Dmodel_Marangoni}b and c), and the size of the depleted zone decreases (to eventually vanish for very small values of $d$) \footnote{This is not obvious from figures \ref{fig:1Dmodel_Marangoni}abc, where the size of the depleted region seems higher for the intermediate value $d/\sigma=5$; however, the $x$-axis represents $x/d$, and $d$ is not the same for those three figures. While $\ell_d$ is decreasing with increasing d, $\ell_d/d$ would present a maximum.}. 
This is also visible in figure \ref{fig:1Dmodel_Marangoni}d for $\sigma/d>0.15$. 
Because $\sigma/d\propto N$, this shows that, as in our experiments, the size of the depleted region is decreasing with the number of swimmers. 

As a very naive application of this model, let us calculate what would be the minimum experimental tank diameter that would allow to host two camphor disks of radius $R=\SI{2.5}{mm}$ with wakes having the same extent as for isolated swimmers. 
From the previous 1D model, non-interacting camphor clouds require $d\ge7\sigma$ in which condition the depleted zone extends over $\ell_d\approx2.8\,\sigma$ (figure \ref{fig:1Dmodel_Marangoni}d).
Overall, the inter-swimmer distance should thus exceed $d\ge2.5\,\ell_d$, which requires that the tank diameter verifies $D\ge d+2\ell_d/2\ge3.5\,\ell_d$. For $R=\SI{2.5}{mm}$, the isolated depleted zone extension is $\ell_d\sim\SI{5.75}{cm}$ from figure \ref{fig:TestOrigin}b or from $A_d^{(1)}$ in figure \ref{fig:deplarea}.
Therefore the minimum diameter for having non-interacting wakes for only two swimmers is $D\sim\SI{20}{cm}$, that is, larger than what we actually have.
This naive application of the model predicts that even for $N=2$, the wakes of each individual swimmer would be smaller than that for a single camphor disk; therefore we would always be in the decaying region, away from the plateau for a tank of the size of our experiment:
this is indeed what we observe in figure \ref{fig:deplarea}a.

Finally, from this model, the physical reason why the depleted region decreases when increasing the number of swimmers should be attributed to the fact that the Marangoni flow $v=-\alpha \partial_z C_t$ is less compressible; 
such an explanation should also hold in our experiment, although the scalings would be different in the 2D case.

\section{Summary and conclusion}

In this article, we have proposed an original experiment of mixing at the free surface of a water tank, with multiple stirrers.  
The stirrers are self-propelled camphor disks that move at the interface of the fluid; 
the particles to be mixed consist in a patch of passive floaters (glass bubbles) initially released at the center of the tank. 
Mixing is achieved thanks to the random motions of $N$ camphor disks, with various $N$ or radii: in a first stage, the decrease of the standard deviation of concentration of glass bubbles is exponential, as for chaotic or turbulent mixing, inline with the power spectra of concentration that exhibit a power law behavior with an exponent close to $-5/3$ in the intermediate spatial frequency range, followed by a second power law with an exponent close to $-1$ at higher frequencies. 
However, the system reaches a stationary state of incomplete mixing, with a final standard deviation of concentration more than twice the mean concentration. 

By averaging the concentration field around a swimmer, we have shown that, in addition to the depleted wake around the swimmer, there is an accumulation front immediately ahead. 
We thus have proposed a very simple model of concentration distribution, with three different values of concentration (an empty wake and an over-concentrated accumulation front around each swimmer, surrounded by a perfectly mixed fluid), that reproduces the levels of unmixing observed through the standard deviation, proving that $C_\mathrm{std}$ is dominated by demixing at large scale. 

In the last section, we have proved experimentally that the depleted area is related to Marangoni effects; then, using rough calculations considering those effects, we have found a good order of magnitude of the size of the depleted area, with a correct scaling for the dependency on the radius of the swimmer. 
Finally, we have proposed a 1D model on Marangoni effects that explains the tendency of the depleted area to decrease when increasing the number of swimmers, even when no crowding effect comes into play. 

Overall, the system reaches a stationary state (although out of equilibrium) of mixing/demixing, where demixing is linked to Marangoni flows, related with compressible effects. 
A striking feature of this study is that, besides demixing, the system develops a "turbulent-like"  concentration spectra, with a large-scale region, an inertial regime at intermediate scale, and a Batchelor regime at small scales: while this is in accordance with the idea that the stirrers are characterized by a "turbulent-like" motion for large enough number of swimmers \cite{bib:bourgoin_etal2020}, this may seem intriguing since the glass-bubbles do not develop such a dynamics; 
this raises the open question of a possible relationship between the spectrum of concentration of a scalar mixed by $N$ moving stirrers and the spatial correlation of the dynamics of those stirrers.

\begin{acknowledgments}
This work was supported by the French research programs ANR-16-CE30-0028, and IDEXLYON of the University of Lyon in the framework of the French program ``Programme Investissements d'Avenir" (ANR-16-IDEX-0005).
\end{acknowledgments}
%
%
%



%
%
%
\newpage
\appendix

\section{Camphor Sublimation\label{ap_sec:camphor}}
One of the camphor characteristics, which has been long invoked to explain its ability to generate long-lived swimmers, is the possibility of constant surface self-cleaning through camphor sublimation into the air. 
The associated flux can be modeled as a first order reaction $J_\mathrm{sub.} = -k C_s$ with $k$ the sublimation rate and $C_s$ the camphor concentration at the surface \cite{Soh_jpcl-2011}; 
in our system the constant has been measured yielding to a rate $k = \SI{6e-7}{\metre\per\second}$ \cite{boniface2019self}.
Assuming sublimation is the dominant mechanism, we assume a rate of sublimation rather than dissolution, and the dissipation term \eqref{eq:CamphorConserv} changes to
\begin{equation}
    Q_s = kC^* A_d\,.
\label{eq:CamphorConserv_alt}
\end{equation}
Assuming production is balanced by sublimation, equation \ref{eq:produc_dissip} now writes
\begin{equation}
    Q_p = Q_s\, ,
\end{equation}
which yields to: 
\begin{equation}
    R_d = 
    \left[
        \frac{\alpha\beta}{\pi k \sqrt{\eta\rho}}
    \right]^{1/2}
    \;\left(\frac{R}{U}\right)^{3/4}.
\label{eq:Lsub_iso}
\end{equation}
Making a numerical estimate using the same typical values as in \eqref{eq:Ldiff_iso}, we predict that $R_d\simeq30R$, therefore now overestimating  the experimental observation $R_d\simeq10R$ (Fig.~\ref{fig:TestOrigin}b). This suggests that unlike mostly assumed, sublimation is not the dominant mechanism for surface cleaning as dissolution  is more efficient to remove camphor from the interface.

Finally, let us note that despite its difference in the orders of magnitude for the depleted zone $R_d$, the scaling obtained with the swimmer size is not significantly affected by the transport mode as \eqref{eq:Lsub_iso} yields $R_d\propto R^{1/2}$ to be compared with $R_d\propto R^{5/9}$ for dissolution-dominated surface cleaning.

\section{Averaged glass bubbles concentration field around a swimmer for all parameters ($N, R$).
\label{appA}}
\begin{figure}[h]
\includegraphics{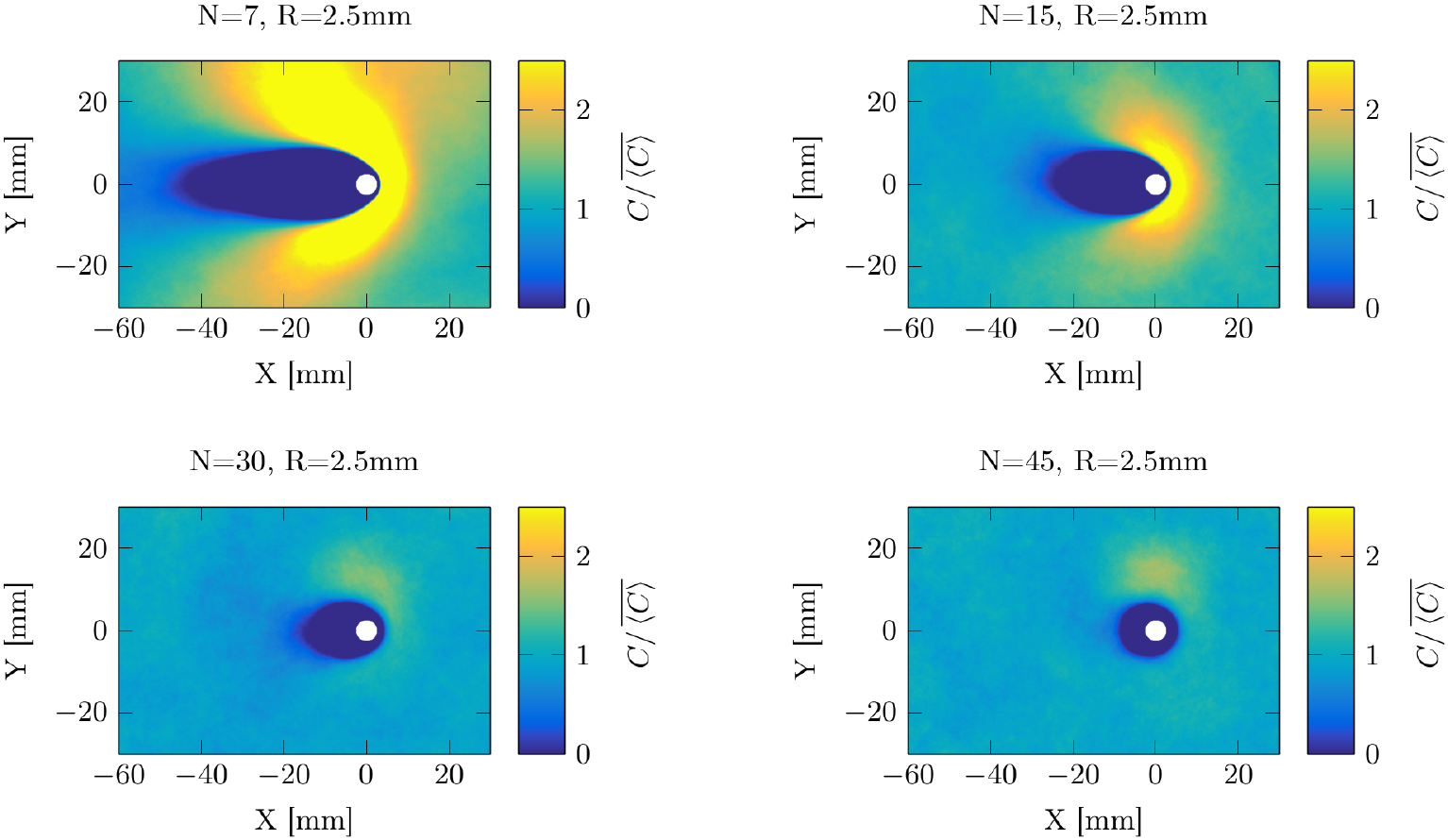}
\caption{Averaged glass bubbles concentration field for different number of swimmers $N$ (fixed radius $R=\SI{2.5}{mm})$.}
\label{fig:meanTrailvarN}
\end{figure}
\begin{figure}[h]
\includegraphics{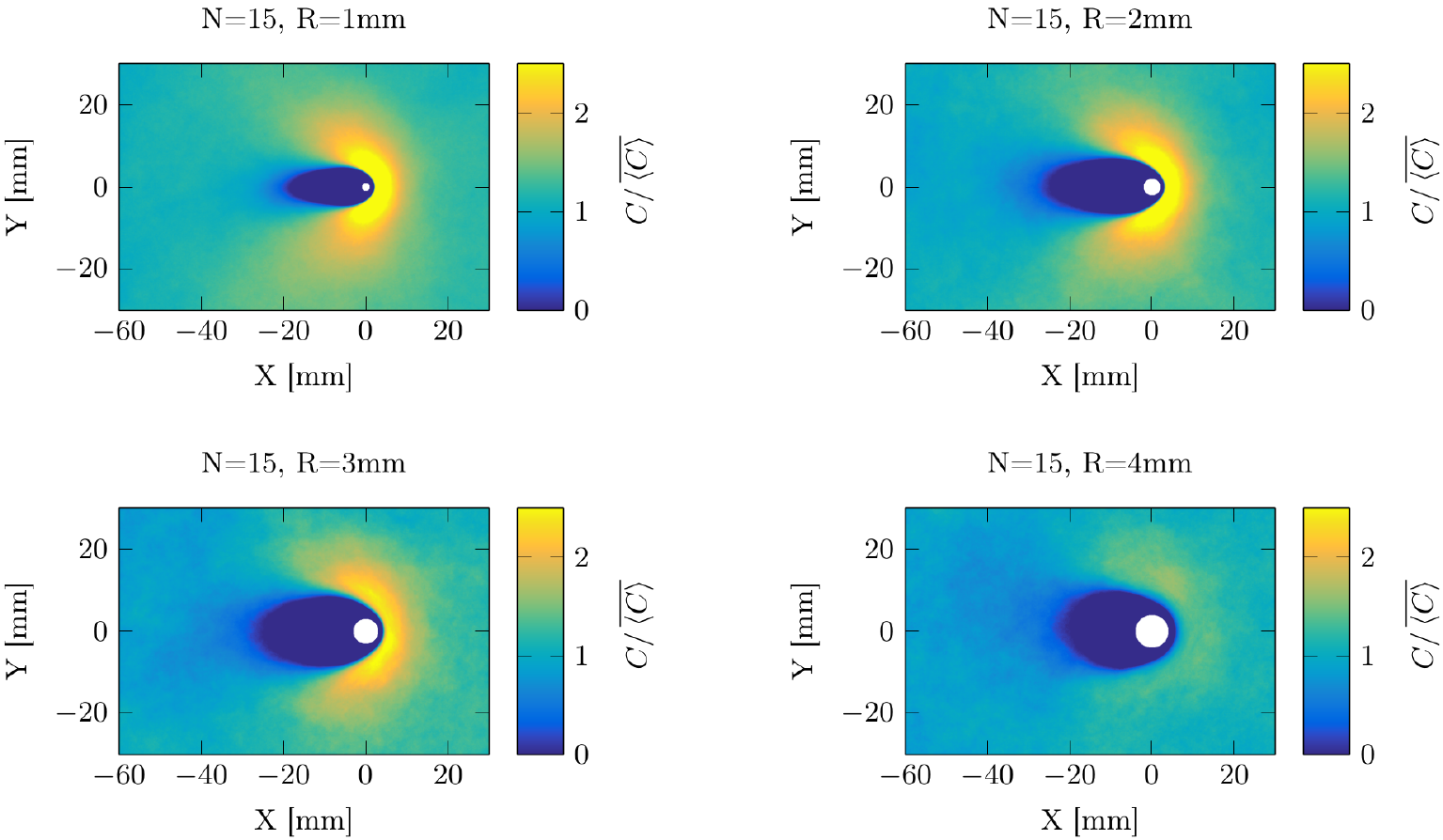}
\caption{Averaged glass bubbles concentration field for different swimmer radii $R$ (fixed number of swimmers $N=15$).}
\label{fig:meanTrailvarR}
\end{figure}

\section{Depleted wake shape
\label{appC}}

\begin{figure}[h]
    \centering
\includegraphics{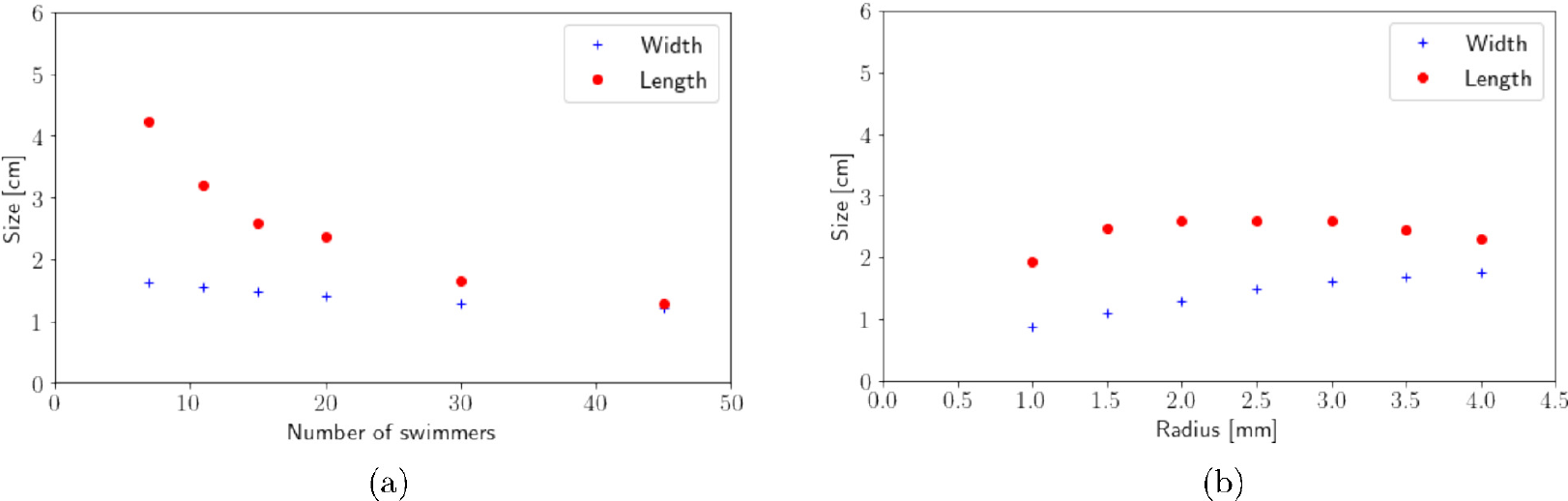}
    \caption{Width $L_\perp$ and length $L$ of the ellipse that best fits the depleted area. (a) for different number of swimmers with equal radius \SI{2.5}{mm}; (b) for $N=15$ swimmers with different radii. 
    }
    \label{fig:size=f(N)}
\end{figure}

As explained in the main text, depletion wakes shown in figures \ref{fig:meanTrailvarN} and \ref{fig:meanTrailvarR} have been fitted by ellipsoids. This provides the overall area already reported, but also the wake shapes -- as defined by their extents respectively perpendicular (width $L_\perp$) and along (length $L$) the swimming direction --.
The figure \ref{fig:size=f(N)} presents these shape parameters ($L_\perp$) and  ($L$) for the various configurations. While the widths $L_\perp$ do not depend much on the number of particles $N$ nor on their radii $R$, the lengths $L$ are reasonably proportional to the velocity of the particle (table \ref{table1}). 

\end{document}